\documentclass[5p,times,twocolumn]{elsarticle}
\usepackage{hyperref} 
\usepackage{ecrc}
\volume{13}
\firstpage{61}
\journalname{Current Opinion in Electrochemistry}
\runauth{B. Horstmann et al.}

\jnltitlelogo{Current Opinion}

\usepackage{amssymb}
\usepackage{amsthm}
\usepackage{color}
\newcommand{\hl}[1]{\textcolor{black}{#1}}

\usepackage[figuresright]{rotating}

\usepackage{amsmath}
\usepackage{bbm}
\usepackage{romannum}
\usepackage[intoc]{nomencl}
\usepackage{enumitem}
\usepackage[version=3]{mhchem}

\begin{document}
\pagenumbering{arabic}
\begin{frontmatter}

\title{Review on Multi-Scale Models of Solid-Electrolyte Interphase Formation}
\author[HIU,DLR]{Birger Horstmann}
\ead{birger.horstmann@dlr.de}
\author[HIU,DLR]{Fabian Single}
\author[HIU,DLR,Ulm]{Arnulf Latz}
\address[HIU]{Helmholtz Institute Ulm (HIU), Helmholtzstra\ss e 11, 89081 Ulm, Germany}
\address[DLR]{German Aerospace Center (DLR), Institute of Engineering Thermodynamics, Pfaffenwaldring 38-40, 70569 Stuttgart, Germany}
\address[Ulm]{Ulm University, Institute of Electrochemistry, Albert-Einstein-Allee 47, 89069 Ulm, Germany}
\begin{abstract}
Electrolyte reduction products form the solid-electrolyte interphase (SEI) on negative electrodes of lithium-ion batteries. Even though this process practically stabilizes the electrode--electrolyte interface, it results in continued capacity-fade limiting lifetime and safety of lithium-ion batteries. Recent atomistic and continuum theories give new insights into the growth of structures and the transport of ions in the SEI. The diffusion of neutral radicals has emerged as a prominent candidate for the long-term growth mechanism, because it predicts the observed potential dependence of SEI growth.   
\\
\\
\emph{Highlights} \\
- Solid-electrolyte interphase passivates negative electrodes in lithium-ion batteries \\
- Recent models elucidate dynamics of solid-electrolyte interphase \\
- Multiple theoretical methods employed: from quantum theory to thermodynamics \\
- Continued capacity fade due to diffusion of neutral radicals \\
\end{abstract}
\end{frontmatter}

\section{Introduction}
\label{sec:intro}
\begin{figure*}[t]
\centering
\includegraphics{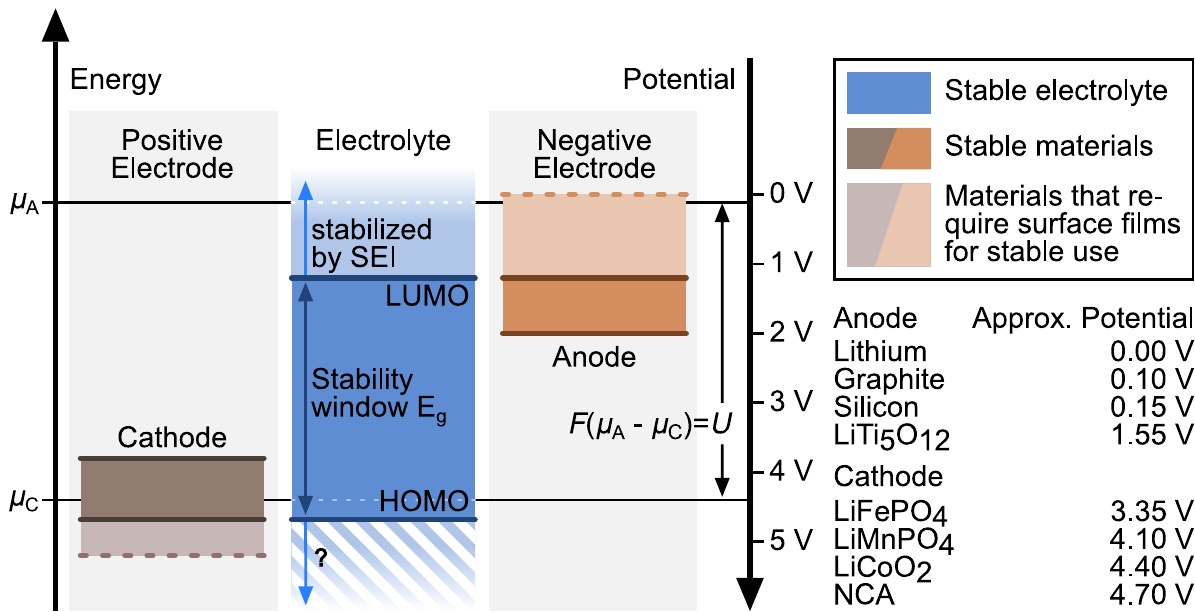}
\caption{Stability of electrode--electrolyte interface in lithium batteries \cite{Goodenough2010}. The positive electrode (left) operates at low energies or high potentials (measured in galvanostatic measurements in EC-PC/LiPF6 solutions at C/20 rates \cite{Martha2011}), whereas the negative one operates at high energies/low potentials \cite{Ohzuku2007}. The stability window is extended by surface films, denoted solid-electrolyte interphase (SEI) on negative electrodes. Stable interfaces are the key for the realization of next-generation low-voltage negative electrodes and high-voltage positive electrodes.}
\label{fig:stability}
\end{figure*}

Standard lithium-ion batteries rely on graphite as negative electrode material even though graphite decomposes the standard electrolytes at their working potentials (see figure \ref{fig:stability}). The decomposition products form the so-called solid-electrolyte interphase (SEI) which is protecting the electrolyte and suppresses further electrolyte reduction  \cite{Paled1979,Aurbach2000}. Nevertheless, lithium transport through the SEI remains possible and is typically not limiting battery performance. The SEI is in the focus of many processes limiting lifetime, performance, and safety of lithium-ion batteries. \hl{It affects the inhomogeneous growth and dissolution of lithium metal \cite{B.HorstmannF.SingleS.HeinT.Schmitt2016,Kushima2017}. Thermal runaway as the main cause for battery failure is promoted by SEI decomposition \cite{Maleki1999,Finegan2015,Abada2016,Tanaka2014}}. The main capacity fade during battery storage stems from the consumption of lithium due to the continued growth of SEI \cite{Keil2016,Keil2017}. During battery cycling, graphite undergoes a notable volume change damaging the SEI and accelerating loss of cycle-able lithium. This volume change is even more pronounced for next-generation high-capacity materials like lithium metal or silicon \cite{Pinson2012}. Generally, the quest for larger battery cell voltages requires improvements in interfacial stability. Thus, SEI modeling contributes to the broad theoretical effort towards rational design of stable electrolytes \hl{\cite{Vatamanu2017,Hoffmann2018,McEldrew2018}}.

\begin{figure*}[t]
\centering
\includegraphics{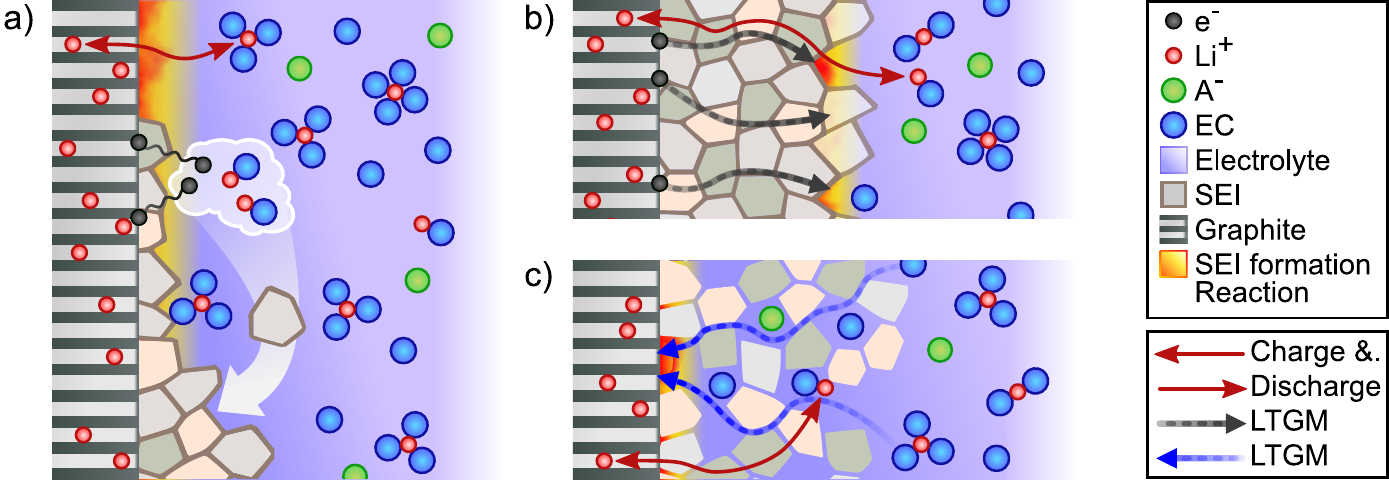}
\caption{
Cross-section through the negative electrode, the SEI, and the electrolyte. Solvent, Li ions and electrons are mobile species and move as indicated by the corresponding arrows. 
(a) Initial SEI formation: Electrons tunnel, electrolyte is reduced and reduction products precipitate as solid film. 
(b) Long-term SEI growth proceeds via a mechanism that transports negative charge to the SEI/electrolyte interface. 
(c) Alternatively, long-term SEI growth is caused by electrolyte diffusing towards the electrode/SEI interface.}
\label{fig:SEImodel}
\end{figure*}

Since 1979 a multitude of experimental SEI research has been performed \cite{Paled1979,Gauthier2015}, recent examples include battery storage at various state-of-charge (SoC) \cite{Keil2016,Keil2017}, differential capacity analysis during cycling \cite{Att2018}, neutron reflectometry \cite{Steinhauer2017}, atomic force microscopy \cite{Kumar2017}, nuclear magnetic resonance \cite{Michan2015}, redox shuttles \cite{Tang2012c,Kranz2017}, fourier-transform infrared spectroscopy \cite{Chapman2017}, and photo-electron spectroscopy \cite{Maibach2016}. As a consequence, there is a general understanding of SEI composition and morphology for few specific systems. The chemical composition of the SEI, however, is diverse and disturbed by trace-amounts of contaminants. Therefore, elucidating SEI behavior requires a careful experimental effort and several key questions about basic SEI mechanisms have yet to be answered (see figure \ref{fig:SEImodel}). Most striking is the fact that the mechanism for \ce{Li+} transport through the SEI is still debated. A dual-layer structure of SEI is typically described with an inner compact layer and a porous polymeric outer layer \cite{Lu2014}, but both the thickness and the formation mechanism of these layers are still debated. 

Under these circumstances, theoretical studies provide important complementary insights into universal principles of SEI chemistry, structure, and dynamics. The diversity of entangled length and time scales governing SEI properties constitutes a fundamental theoretical challenge. One should, for example, distinguish between the process of initial SEI formation in hours and days and the continued SEI growth in months and years. \hl{On the one hand, SEI chemistry is governed by reactions between individual atoms and molecules. On the other hand, molecular environments influence reaction pathways and transport through the SEI determines the availability of reactants.} Therefore, we begin with a brief summary of results from atomistic theories based on quantum physics as they are prerequisites for multi-scale models on larger scales. 

This review, however, focuses on recent continuum models based on thermodynamics. These meso-scale models discuss emergent phenomena of SEI formation, particularly, the origin of continued SEI growth. SEI thickness is experimentally observed to grow with the square-root of time during storage under controlled lab conditions. Therefore, a transport process seems to limit SEI growth after sufficiently long times. Continuum models evaluate various long-term growth mechanisms  
\begin{enumerate}[itemsep=0ex,label=\alph*)]
\item \label{mech1} Electron tunneling \cite{Tang2012,Li2015}
\item \label{mech2} Diffusion of solvent/salt molecules/anions \cite{Ploehn2004a,Pinson2012,Tang2011,Tang2012b,Tang2012,Roder2016,Single2016a,Single2017,Tahmasbi2017a,Hao2017,Single2018}
\item \label{mech3} Electron conduction or diffusion \cite{Broussely2001,Christensen2004,Colclasure2011,Tang2012,Single2016a,Single2017,Roder2017,Single2018,Das2018}
\item \label{mech4} Diffusion of neutral radicals such as lithium interstitials \cite{Shi2012,Soto2015,Single2017,Single2018,Das2018}
\end{enumerate}
\hl{Most models describe the ideal square-root-of-time dependence of capacity fade. Electron tunneling, however, predicts capacity fade with the logarithm of time as discussed below. Some articles model battery operation and analyze linear growth regimes.} In this review, we highlight models that predict additional observable properties, i.e., morphology of SEI \cite{Tang2012,Single2016a,Single2017}, explain additional dependencies, i.e., potential dependence of SEI growth \hl{\cite{Single2018,Das2018}}, or analyze non-ideal settings, i.e., SEI growth during cycling \cite{Das2018}. These allow the experimental validation of proposed growth mechanisms.

\section{Atomistic theories and initial SEI growth}
Atomistic simulation methods address elementary reaction and transport processes in the SEI. Energies of atom configurations in electrolytes are probed with quantum chemistry and density functional theory (DFT). The resulting energy landscape determines forces between atoms and reaction probabilities. The collective dynamics of molecules and atoms can then be calculated with molecular dynamics simulations (MD). \hl{In this section}, we give a brief outline of results from atomistic simulations, but refer to recent reviews for further details \cite{Soto2016,Li2016,Borodin2017,Wang2018}. 

Borodin et al. highlight general challenges for calculations of electrolyte stability \cite{Borodin2015}. Solvent and solutes interact so strongly that calculations on individual molecules are inaccurate. This necessitates large simulation domains and optimized molecular geometries. The diverse SEI chemistry imposes further challenges. It has been shown with DFT and ab-initio MD that salt anion \cite{MartinezDeLaHoz2015,Chapman2017} and electrode voltage \cite{Leung2015} affect electrolyte stability and chemical SEI composition. Nevertheless, recent calculations provide further insights into preferred reduction pathways in conventional lithium battery electrolytes comprising a mixture of ethylene carbonate (EC) and linear carbonates, e.g. dimethyl carbonate (DMC). In agreement with experimental observations, it is rationalized that EC is preferentially reduced because EC has a higher reduction potential than DMC \cite{Borodin2015}, EC preferential adsorbs on the SEI surface \cite{Borodin2014}, and \ce{Li+} prefers EC in its inner solvation shell \cite{Borodin2016}. 

Atomistic theories alone can only address the initial stages of SEI formation because of limits in simulated space and time \cite{Bertolini2018,Takenaka2014}. \hl{Electron tunneling allows the transport of electrons through $2$-$3\text{ nm}$ thin SEI layers \cite{Lin2016}, while SEI thickness quickly exceeds $10 \text{ nm}$ \cite{Nie2013,Lu2014}. This suggests that electron tunneling plays a role only in the initial part of first-cycle SEI growth (see fig. \ref{fig:SEImodel}a)}. \ce{Li2O} is predicted to form the innermost SEI layer on the electrode surface at low potentials \cite{Leung2016}. Furthermore, nucleation and precipitation play an important role in the initial SEI formation \cite{Ushirogata2015}. 

Furthermore, the mechanism for \ce{Li+} transport through the SEI is analyzed with atomistic calculations. For the inner inorganic layer, different lattice diffusion mechanisms in crystalline \ce{LiF}, \ce{Li2O} and \ce{Li2CO3} are compared \cite{Shi2012,Shi2013,Soto2017,Benitez2017}. Alternatively, \ce{Li+} is proposed to diffuse along interfaces between these crystalline phases \cite{Zhang2016}. For the outer organic layer, MD determines diffusion constants of \ce{Li+} through ordered and disordered \ce{LiEDC} \cite{Bedrov2017}. 
Besides transport of \ce{Li+}, atomistic theories discuss mechanisms for electron transport in the SEI. We highlight the recent proposals of diffusion of neutral lithium interstitials through the crystalline inner layer \cite{Shi2012,Shi2013} and radical diffusion through the polymeric and amorphous outer layer \cite{Soto2015}. These mechanisms lay the foundation for novel models of continued SEI growth (see Sec. \ref{sec:interstitials}).

\section{Continuum models and long-term SEI growth}
In 2001, Broussely et al. recorded the lifetime of lithium-ion batteries and observed a continued capacity fade due to SEI growth \cite{Broussely2001}.  \hl{Assuming transport-limited SEI growth and neglecting the electrochemical details, they derive a rate equation for SEI thickness evolution. This prototype model demonstrates that sluggish electron transport through the SEI would explain the observed square-root-of-time behavior of capacity fade.} Subsequent modeling studies elaborate on this model and present various long-term growth mechanisms (LTGM) \cite{Christensen2004,Ploehn2004a}. On the one hand, the coupled diffusion and/or migration of negative charges, e.g. electron conduction, from the graphite/SEI interface to the SEI/electrolyte interface predicts the observed SEI growth \cite{Christensen2004}. (see fig. \ref{fig:SEImodel}b). On the other hand, the diffusion of electrolyte constituents, e.g. solvent molecules, from the SEI/electrolyte to the graphite/SEI interface agrees equally well with SEI thickness evolution \cite{Ploehn2004a} (see fig. \ref{fig:SEImodel}c). Note that the core mathematical description of SEI thickness is equivalent for both LTGMs. To conclude, continuum models should predict measurable properties beyond SEI thickness in order to determine the LTGM.

A coupled multi-species model found a minor influence of cycling on SEI thickness \cite{Colclasure2011}. Cell-level models conclude that SEI thickness varies little in a porous electrode \cite{Pinson2012,Tahmasbi2017a}. Pinson and Bazant extend their SEI model and describe the rapid capacity decrease during cycling \cite{Pinson2012}. Because the drastic volume change of silicon electrodes stresses the SEI, a constant rate of SEI cracking is assumed. Therefore, SEI thickness deviates from the square-root-of-time law and grows linearly in time, as observed experimentally on silicon anodes. Coupled models of continuum mechanics and electrochemistry begin to take a closer look at SEI fracture \cite{Tanaka2018}. The combination of continuum simulations of transport with stochastic Monte Carlo simulations of reduction reactions gives further microscopic insights, but has not yet lead to new macroscopic predictions \cite{Roder2016,Hao2017,Roder2017}. 

Some articles analyze the role of electron tunneling for long-term SEI growth. Because capacity fade would grow with the logarithm of SEI thickness, Tang et al. discard electron tunneling as possible LTGM \cite{Tang2012}. Nevertheless, a model based on electron tunneling has recently been fitted to capacity fade experiments \cite{Li2015}. In this model, the growth of the outer SEI layer is controlled by electron tunneling through \hl{an approximately $3\text{ nm}$ thin} inner layer. The ratio of growth of the inner versus the outer layer is determined by model assumption such that the inner layer does not even grow a single mono-layer during the long-term experiment. \hl{We note that the capacity fade experiments discussed in Ref. \cite{Li2015} can be fitted equally well with square-root-of-time-growth as with logarithm-of-time-growth. This demonstrates that time dependence of capacity fade as single metric cannot prove the correctness of a SEI growth model.}

In a comprehensive experimental and theoretical approach, Tang et al. study SEI formation and redox shuttles at negative electrodes \cite{Tang2011,Tang2012b,Tang2012}. A dual-layer SEI with a compact inner and a porous outer layer is modeled with a volume-averaged transport theory by introducing a constant porosity in each layer $\varepsilon$. They aim at determining the LTGM by comparing experiments with different models, each based on a single rate-limiting mechanism \cite{Tang2012}. The square-root-of-time behavior restricts possible LTGMs to transport mechanisms, e.g., solvent diffusion and electron conduction. Solvent diffusion fails to explain the observed dependence of growth rate on electrode potential. Electron conduction fails to explain the involvement of convection in SEI growth. They finally conclude that another form of charge transport must be rate-limiting.

\begin{figure*}[t]
	\centering
		\includegraphics[width=\textwidth]{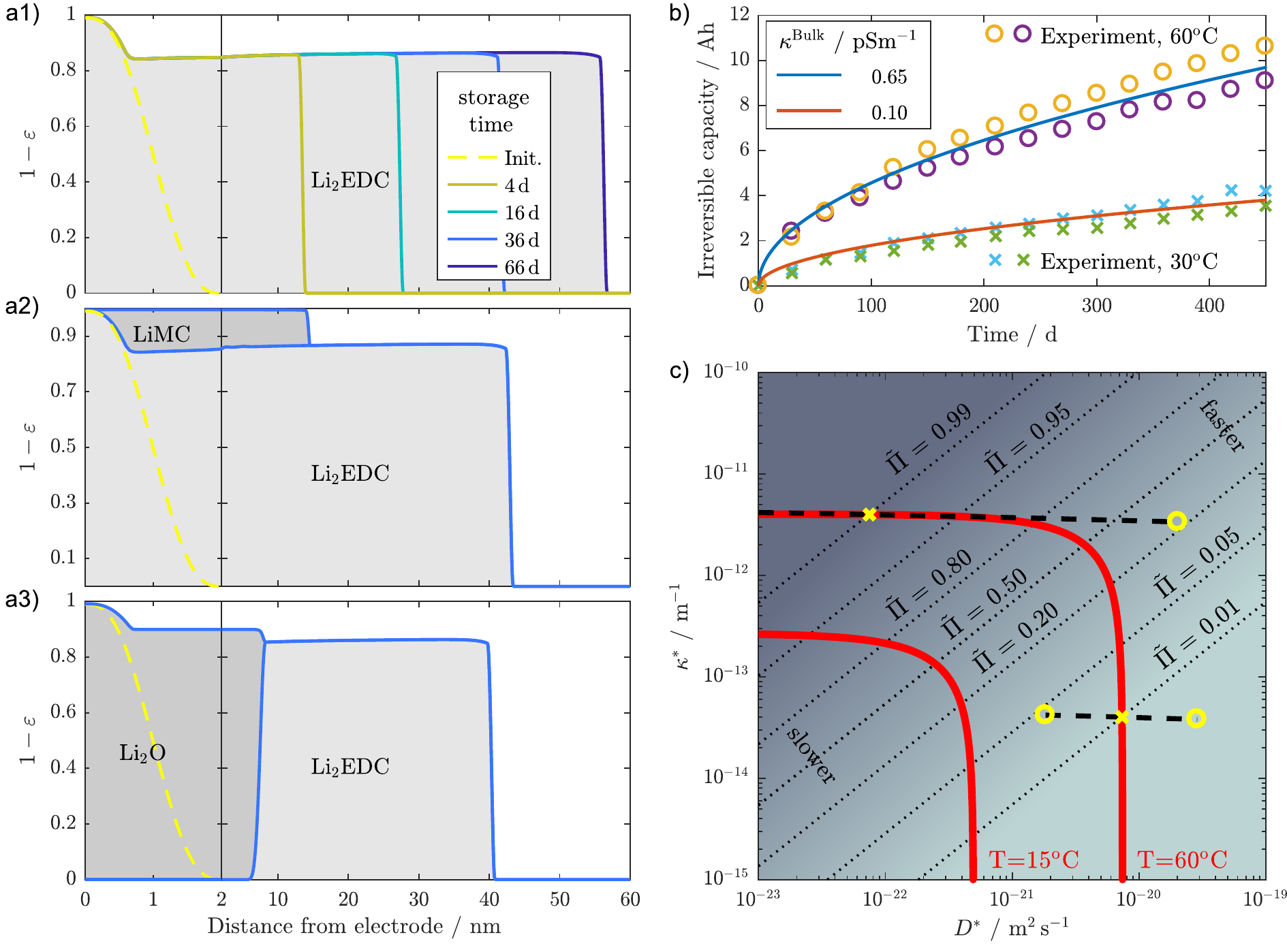}
	\caption{Simulation of long-term SEI growth. (a) Time evolution of the SEI volume fraction for (a1) a single-layer SEI from reduction of \ce{EC} to \ce{Li2EDC}, (a2) a dual-layer SEI due to reduction of co-solvent \ce{DMC}, and (a3) a dual-layer SEI due to conversion of \ce{Li2EDC}. (b) Simulated capacity fade according to the electron conduction mechanism with conductivity $\kappa^\text{Bulk}$ (lines) compared to experimental data (circles and crosses) \cite{Broussely2001} close to the upper yellow cross in (c). (c) Relative position of the reaction interface $\tilde \Pi$ depending on the effective transport parameters $D^*$ and $\kappa^*$. The red lines show parameter sets with identical SEI growth rates. The \hl{dashed} black lines end in yellow circles where the formation rate is double (right) or half (left) of the original growth rate. Reproduced from Single et al. \cite{Single2017}.}
\label{fig:sei_growth_single}
\end{figure*}

The recent models of Single et al. take into account two counter-propagating transport processes, i.e., motion of charges from the electrode to the electrolyte and motion of solvent molecules from the electrolyte to the electrode \cite{Single2016a,Single2017}. This allows to predict not only SEI thickness, but also SEI porosity $\varepsilon(x,t)$. A volume-averaged transport model determines the spatially-resolved dynamics of solvent, electric potential, and SEI porosity. Modeling convection of solid SEI facilitates simulating reduction reactions inside the SEI. 

A single-layer SEI comes out if solvent \ce{EC} is reduced to \ce{Li2EDC} and co-solvent \ce{DMC} is inert \cite{Single2016a,Single2017}. A typical evolution of SEI volume fraction is depicted in fig. \ref{fig:sei_growth_single}a1. It is found that SEI growth is limited by electron transport and that SEI predominantly grows at the electrolyte/SEI interface \cite{Single2016a} ($\tilde\Pi \approx1$ in fig. \ref{fig:sei_growth_single}c). Therefore, SEI thickness grows like the square-root-of-time in agreement with capacity fade experiments (see fig. \ref{fig:sei_growth_single}b). The predicted SEI porosity is almost constant and approaches a stability point $\varepsilon^*$ determined by electrolyte transport properties. The transition from electron conduction to solvent diffusion as LTGM is studied by imposing large SEI porosity and taking into account solid convection \cite{Single2017}. If solvent diffusion is rate-limiting, the reaction zone moves to the electrode/SEI interface ($\tilde\Pi\approx 0$ in fig. \ref{fig:sei_growth_single}c) and \hl{significant} fluctuations in SEI thickness are predicted. 

Additional SEI formation reactions lead to a dual-layer SEI \cite{Single2017}. If reduction of co-solvent \ce{DMC} or primary SEI compound \ce{Li2EDC} is considered, low potentials favor the second reduction near the electrode and a compact, non-porous, inner layer is formed (see fig. \ref{fig:sei_growth_single}a2,a3). The ratio between the thickness of inner and outer layer is determined by electrode potential and material parameters. Simulations illustrate that this stationary thickness ratio is quickly re-attained after the SEI is disturbed. Most importantly, SEI thickness and capacity fade grow with the square-root of time for dual-layer morphologies, as well.

\section{Multi-scale models of electron leakage via neutral radicals}
\label{sec:interstitials}
\begin{figure}[t]
	\centering
		\includegraphics[width=\columnwidth]{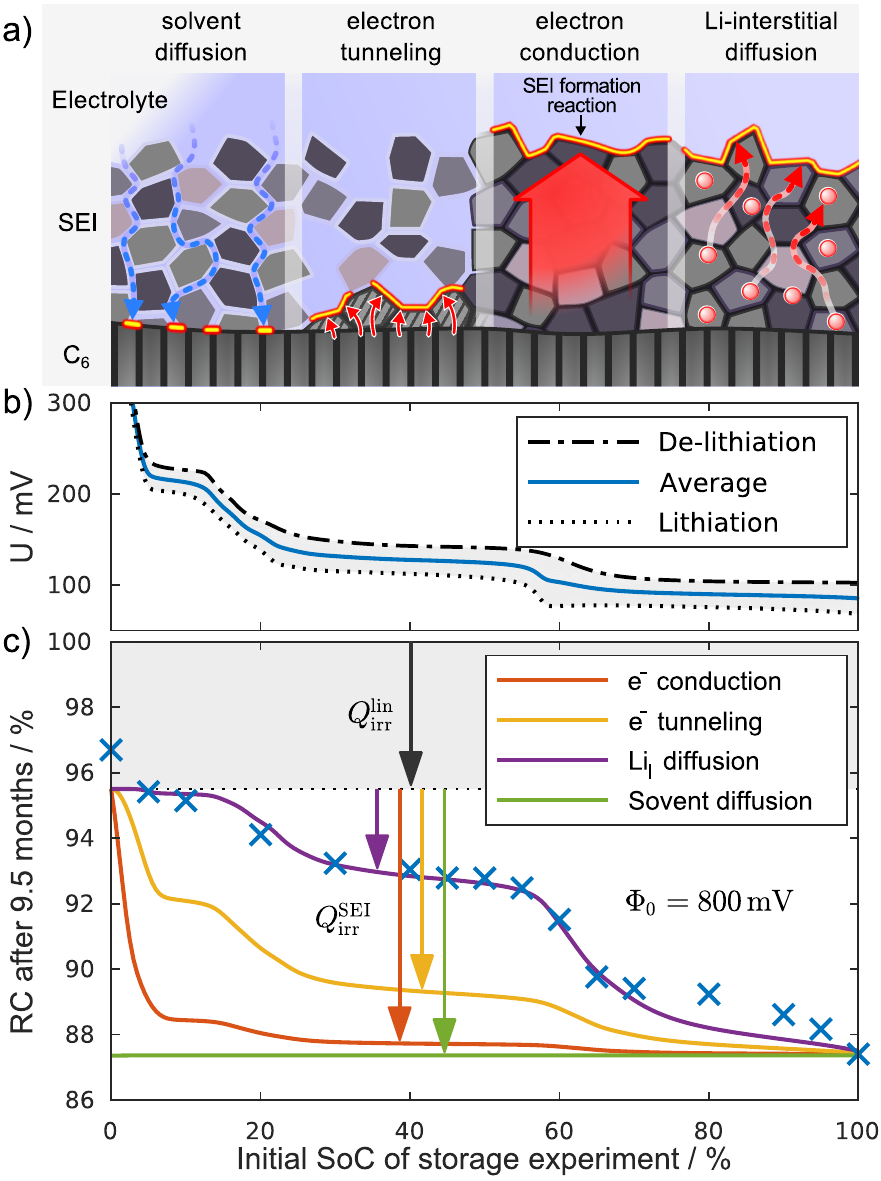}
	\caption{(a) Schematic of all four candidates for transport mechanisms: Solvent diffusion through small SEI pores, electron tunneling through a thin and dense inner SEI layer, electron conduction through the SEI, diffusion of neutral Li-interstitials through the SEI. The SEI formation reaction takes place at different interfaces depending on the mechanism, marked yellow/red.
	(b) Open circuit voltage of the negative electrode gained by averaging the lithiation and delithiation voltages (half cell, cycled at C/20). (c) Experimentally obtained relative capacity after 9.5 months of storage (crosses) compared to that predicted by four different long-term growth mechanisms (lines). Reproduced from Single et al. \cite{Single2018}.}
	\label{fig:scheme_interstitial_diffusion}
\end{figure}

Based on atomistic theories, Shi et al. and Soto et al. propose diffusion of neutral radicals as \hl{an} alternative mechanism for charge transport from the electrode through the SEI into the electrolyte \cite{Shi2012,Soto2015}. In the case of inorganic SEI, lithium ions take up an electron at the electrode/SEI interface, diffuse as neutral lithium interstitials through the SEI, and release an electron at the SEI/electrolyte interface \cite{Shi2012}. In the porous organic SEI, radicals formed by electrolyte reduction can act as electron carrier \cite{Soto2015}. Single et al. take up this result and develop a continuum model based on diffusion of neutral radicals \cite{Single2017}. SEI profiles simulated with this mechanism share the same features as those described above for electron conduction. \hl{Recent continuum models highlight the unique exponential dependence of SEI growth rate on electrode potential for this mechanism \cite{Single2018,Das2018}.}

\hl{The first such model by Single et al. points out that the concentration of radicals at the electrode is determined by its electric potential \cite{Single2018}.} They compare the predictions of different LTGMs with capacity fade experiments for various graphite potentials and state-of-charges (SoC) \cite{Keil2017}. Simple theories based on the four LTGMs enlisted above are created: electrolyte diffusion, electron tunneling, electron conduction, and lithium interstitial diffusion. As summarized in fig. \ref{fig:scheme_interstitial_diffusion}, solvent diffusion does not reproduce any SoC dependence. The SoC dependence of electron conduction and electron tunneling does not agree with the experiment for any reasonable choice of parameters. Only a mechanism such as neutral lithium interstitial diffusion results in a promising agreement with the experiment and remains a candidate for the LTGM.

Recently, Das et al. extend this model and couple SEI growth with lithium-ion transport through the SEI \cite{Das2018}. Based on atomistic theories \cite{Shi2012}, they assume diffusion of lithium ions on interstitial sites and electron conduction on this sparse network of lithium-ion interstitials. Note that an electron bound to a lithium-ion interstitial constitutes the aforementioned neutral lithium interstitial. As a consequence, the concentration of lithium ions determines electron conductivity. This model can explain recent differential capacity measurements that SEI grows only during lithiation, but not during delithiation \cite{Att2018}. 

\section{Conclusions}
\label{sec:conclusions}
In this short review, we summarize recent theoretical studies of SEI structure and formation. A multi-scale approach is necessary to elucidate the broad range of SEI properties from chemical composition to mechanical structure. Predictions of atomistic theories converge towards a clear SEI chemistry for standard carbonate-based electrolytes, but the relevance of transport mechanisms remains debated. 

Continuum models build on recent findings and demonstrate macroscopically observable consequences of microscopic material behavior. Understanding SEI formation is a key goal. We distinguish between formation of initial SEI and long-term SEI growth. Recent simulations explain the SEI dual-layer structure. Diffusion of neutral radicals leads to the observed potential dependence of long-term growth mechanisms. Coupling this mechanism with lithium-ion diffusion predicts an observed asymmetry in SEI growth during cycling. 

\section{Acknowledgement}
This work is supported by the German Federal Ministry of Education and Research (BMBF) in the project Li-EcoSafe (03X4636A). Further support was provided by the bwHPC initiative and the bwHPCC5 project through associated compute services of the JUSTUS HPC facility at the University of Ulm. \hl{This work contributes to the research performed at CELEST (Center for Electrochemical Energy Storage Ulm-Karlsruhe).}

\section*{Highlighted References}
\begin{small}
\noindent [28] ** M. Tang, S. Lu, J. Newman, J. Electrochem. Soc. 2012, 159, A1775-A1785. ** In a comprehensive theoretical and experimental approach, Tang et al. compare various transport mechanisms as possible long-term growth mechanism.

\noindent [35] * F. Single, B. Horstmann, A. Latz, J. Electrochem. Soc. 2017, 164, E3132-E3145. * This paper models not only SEI thickness, but also SEI porosity and SEI dual-layer structure. The transition from electron conduction to solvent diffusion as long-term growth mechanism is studied.

\noindent [38] ** F. Single, A. Latz, B. Horstmann, ChemSusChem 2018, 11, 1950-1955. ** In this article, Single et al. compare the SoC dependence of various long-term growth mechanisms with storage experiments and present the first indirect experimental evidence for neutral interstitial diffusion.

\noindent [43] * S. Das, P. M. Attia, W. C. Chueh, M. Z. Bazant, arXiv 2018, to be submitted. * By coupling neutral lithium interstitials with lithium-ion interstitials, the authors explain recent experiments showing asymmetric SEI growth during (de-)intercalation.

\noindent [44] ** S. Shi, P. Lu, Z. Liu, Y. Qi, L. G. Hector, H. Li, S. J. Harris, J. Am. Chem. Soc. 2012, 134, 15476-15487. ** This article presents atomistic calculations of transport mechanisms in inorganic SEI material. For the first time, it proposes diffusion of neutral lithium interstitials as SEI growth mechanism.

\noindent [45] * F. A. Soto, Y. Ma, J. M. Martinez De La Hoz, J. M. Seminario, P. B. Balbuena, Chem. Mater. 2015, 27, 7990-8000. * This article proposes the formation of organic radicals which allow electron leakage through the outer organic SEI layer.

\noindent [50] * O. Borodin, M. Olguin, C. E. Spear, K. W. Leiter, J. Knap, Nanotechnology 2015, 26, 354003. * Borodin et al. highlight general challenges for atomistic calculations of electrolyte stability.
\end{small}

\bibliographystyle{elsarticle-num}
\bibliography{library}

\end{document}